\documentclass[12pt]{iopart}

\usepackage{graphicx}
\usepackage[usenames,dvipsnames]{xcolor}
\usepackage{amssymb, amsthm}
\usepackage{iopams}
\usepackage{float}
\usepackage{cite}
\usepackage[colorlinks]{hyperref}

\newcommand{\gE}{g_\mathrm{E}}
\newcommand{\muB}{\mu_\mathrm{B}}
\newcommand{\HS}{\emph{half shell}}
\newcommand{\FS}{\emph{full shell}}

\begin{document}

\title[Shell-shaped condensates with gravitational sag]{Shell-shaped condensates with gravitational sag: contact and dipolar interactions}

\author{Maria Arazo$^{1,2}$, Ricardo Mayol$^{1,2}$ and Montserrat Guilleumas$^{1,2}$}

\address{$^1$ Departament de F\'isica Qu\`antica i Astrof\'isica, Facultat de F\'isica, Universitat de Barcelona, Mart\'i i Franqu\`es 1, 08028 Barcelona, Spain}

\address{$^2$ Institut de Ci\`encies del Cosmos, Universitat de Barcelona, Mart\'i i Franqu\`es 1, 08028 Barcelona, Spain}

\ead{maria.arazo@fqa.ub.edu}

\begin{abstract}
We investigate Bose-Einstein condensates in bubble trap potentials in the presence of a small gravity. In particular, we focus on thin shells and study both contact and dipolar interacting condensates. We first analyze the effects of the anisotropic nature of the dipolar interactions, which already appear in the absence of gravity and are enhanced when the polarization axis of the dipoles and the gravity are slightly misaligned. Then, in the small gravity context, we investigate the dynamics of small oscillations of these thin, shell-shaped condensates triggered either by an instantaneous tilting of the gravity direction or by a sudden change of the gravity strength.
This system could be a preliminary stage for realizing a gravity sensor in space laboratories.
\end{abstract}

\maketitle


\section{Introduction}

The recent progress in microgravity experiments with Bose-Einstein condensates (BECs) \cite{vanZoest2010,Becker2018,Condon2019,Muntinga2013} and the development of new exotic confining potentials \cite{Zobay2004,Garraway2016,White2006,Zobay2001,Perrin2017,Harte2018} have fostered a novel field of research on shell-shaped BECs.
These hollow condensates were first realized in 2004 \cite{Colombe2004} and are at present under investigation in the NASA Cold Atom Laboratory (CAL) on the International Space Station \cite{Lundblad2019,Elliott2018,Aveline2020}.
Due to Earth’s gravity, the atoms sag to the bottom of the trap destroying shell-shaped BECs.
Thus, microgravity conditions---in which gravity is small enough to be neglected---ensure that these condensates are realizable in experiments.
The observation of a shell-shaped BEC at CAL \cite{Lundblad2019} has spiked the interest in such condensates under microgravity conditions \cite{Meister2019,Tononi2019,Tononi2020}.
In this paper, however, we are interested in the effect of a small gravity in shell-shaped BECs. Hence, to ensure that shells are realizable, we consider gravities larger than microgravity to study its effect but still some orders of magnitude smaller than the terrestrial gravity.

BECs in shell-shaped potentials open the possibility to investigate condensation and superfluidity phenomena in new topologies:
collective modes \cite{Lannert2007,Sun2018,Merloti2013},
self-interference effects \cite{Tononi2020},
thermodynamics of shells and curved manifolds \cite{Bereta2019,Moller2020},
quantized vortices \cite{Tononi2019, Padavic2020, Bereta2021},
topological transitions in curved systems \cite{Tononi2021},
the dimensional reduction to a ring-shaped condensate \cite{Guo2021},
and the transition from filled to hollow condensates \cite{Padavic2017,Rhyno2021}, among others.
Theoretical work has focused mainly on shell-shaped BECs with contact-interacting atoms rather than with atoms that possess a non-negligible dipolar moment. The latter situation has been examined in the limit of a thin shell \cite{Diniz2020} and under rotation \cite{Adhikari2012}.

While contact interactions are short-range and isotropic, the interaction between particles with a dipolar moment presents a long-range and anisotropic character \cite{Lahaye2009}. This intrinsic feature of dipolar BECs makes these systems especially sensitive to the shape of the trapping potential. Besides, the existence of a privileged direction defined by the dipole polarization endows dipolar BECs with an interesting sensitivity to small changes in orientation, such as perturbations of gravity.

In this work, we investigate the effects of the dipolar interaction in thin shell-shaped condensates. Moreover, since experiments in microgravity conditions or at CAL facilities might suffer gravity perturbations, we study the dynamics of small oscillations in small-gravity conditions, which could yield to identify small changes in the direction or magnitude of the gravity.

The paper is structured as follows. Section ~\ref{sec:theory} introduces the shell-shaped potential and the theoretical framework. In section~\ref{sec:ground}, we analyze the ground state configurations---both for contact and dipolar interacting BECs---in the presence of gravitational sag. Then, we discuss two cases: when the gravity is parallel with the $z$-axis---the polarization direction in dipolar BECs---and when it is slightly misaligned. Section~\ref{sec:dynamics} explores the dynamics of small oscillations triggered by a tiny variation in the gravity direction or its strength. Lastly, we summarize our results and provide future perspectives in section~\ref{sec:conclusions}.


\section{Theoretical framework}\label{sec:theory}

We consider $N$ dilute and weakly-interacting dipolar bosons at zero temperature confined in a shell-shaped potential $V_{\mathrm{ext}}(\vec{r})$.
In the mean-field framework, the Gross-Pitaevskii equation (GPE) provides a good description of a weakly interacting dipolar BEC:
\begin{equation}\label{eq:tdgpe}
    \left[
    - \frac{\hbar^2}{2m}\vec{\nabla}^2 \!
    + \!V_{\mathrm{ext}}(\vec{r}) \!
    + g |\Psi(\vec{r},t) |^2 \!+ \!V_{\mathrm{dd}}(\vec{r})
     \right] \!\! \Psi(\vec{r},t) = i \hbar \, \frac{\partial  \Psi(\vec{r},t)}{\partial t}\,,
     \label{GPE}
\end{equation}
where $\Psi(\vec{r},t)$ is the condensate wave function normalized to the total number of particles $N$.
The atom-atom mean-field interaction is characterized by the contact-interacting potential with coupling constant $g= 4 \pi \hbar^2 a_\mathrm{s}/m$, where $a_\mathrm{s}$ is the s-wave scattering length and $m$ the atomic mass, and the dipolar interaction
 $V_{\mathrm{dd}}(\vec{r}) =  \int|\Psi|^2 v_{\mathrm{dd}}(\vec{r}-\vec{r}^{\,\prime})\,\mathrm{d}\vec{r}^{\,\prime}$.
The dipolar interaction potential for a polarised sample of particles with dipolar moment $\vec{\mu}$ oriented along the $z$-axis is
    \begin{equation}
    v_{\mathrm{dd}}(\vec{r}-\vec{r}^{\,\prime}) = \frac{C_{\mathrm{dd}}}{4\pi}\frac{1-3\cos^2\theta}{|\vec{r}-\vec{r}^{\,\prime}|^3} \,,
    \end{equation}
where $|\vec{r}-\vec{r}^{\,\prime}|$ is the relative distance between particles, $\theta$ is the angle between $\vec{r}-\vec{r}^{\,\prime}$ and the direction of polarization and $C_{\mathrm{dd}}$ is $\mu_0\mu^2$ ($d^2/\epsilon_0$) for a magnetic (electric) dipole moment.

Analogously to contact interactions---which are characterized by the s-wave scattering length---, a dipolar effective length can be introduced for dipole-dipole interactions, $a_{\mathrm{dd}}=C_{\mathrm{dd}}m/(12\pi\hbar^2)$. Then, the relative strength of both interactions is defined as the ratio of these two effective lengths, $\epsilon_{\mathrm{dd}}=a_{\mathrm{dd}}/a_\mathrm{s}$, which in the case of magnetic moments is
\begin{equation}
    \epsilon_{\mathrm{dd}} = \frac{\mu_0\mu^2m}{12\pi\hbar^2a_\mathrm{s}} \,.
\end{equation}

Shell-shaped BECs have been experimentally realized by employing time-dependent, radio-frequency induced adiabatic potentials within a conventional magnetic trap \cite{Zobay2004}. In the thin-shell limit, where the thickness of the shell is small compared to its radius, these bubble trap potentials can be approximated by a radially shifted harmonic trap \cite{Sun2018,Padavic2017}:
    \begin{equation}
    \label{eq-pot}
        V_{\mathrm{ext}}(\vec{r}) = \frac{1}{2}m\omega^2 \left( r - r_0 \right)^2 .
    \end{equation}
This potential defines a spherically symmetric shell of radius $r_0$, with $\omega\equiv \omega_x=\omega_y=\omega_z$ and $r^2\equiv x^2+y^2+z^2$.


\section{Ground states}\label{sec:ground}

From now on, we consider a typical BEC in the mean-field regime: $N=10^4$ atoms of ${}^{164}$Dy polarized along the $z$-axis with magnetic dipolar moment $\mu=10 \, \muB$, scattering length $a_\mathrm{s}=120 \, a_0$, and mass $m=164~\mathrm{amu}$. In this case, the relative strength of the interactions is $\epsilon_{\mathrm{dd}}=1.11$. 

We start by characterizing the ground state of the system with and without gravitational sag. In the spherical-shell geometry described before, we obtain the shell-shaped ground state wave function by solving the time-independent GPE with the imaginary-time propagation method in 3D.
The dipolar term transforms the GPE into a more complicated equation. However, one can evaluate the dipolar interaction integral, $V_{\mathrm{dd}}(\vec{r})$, employing Fourier transform techniques---see \cite{Abad2009} and references therein. In particular, we use the FFTW package \cite{FFTW}.
In all the results we present, the numerical grid is a 3D box of 16~$\mu$m $\times$ 16~$\mu$m $\times$ 16~$\mu$m.


\subsection{Dipole-dipole interactions}

The dipolar interaction deforms the ground state density to minimize the energy of the system. This effect is a consequence of the anisotropic character of the dipolar interactions and depends on the specific trapping potential. It was observed already in the first dipolar condensates as the appearance of new structured biconcave ground states for some particular values of the strength of the dipolar interactions and the harmonic trap anisotropy \cite{Lahaye2009}. Afterward, this feature was proposed to generate a self-induced bosonic Josephson junction in a toroidally confined dipolar condensate \cite{Abad2011,Abad2015}.

\begin{figure}[t]
\centering
    \includegraphics[width = 0.51\linewidth]{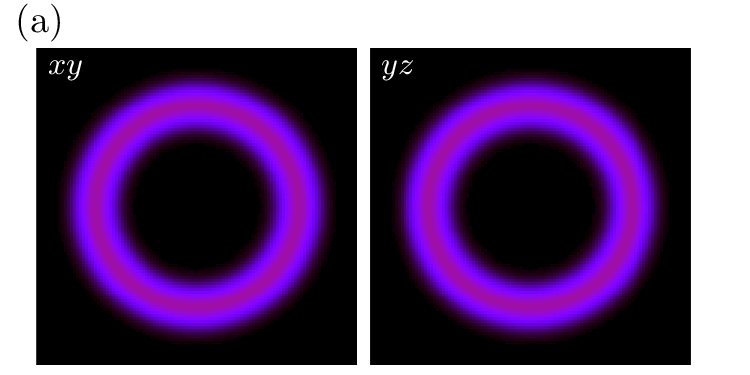}\hspace{-1.1em}
    \includegraphics[width = 0.51\linewidth]{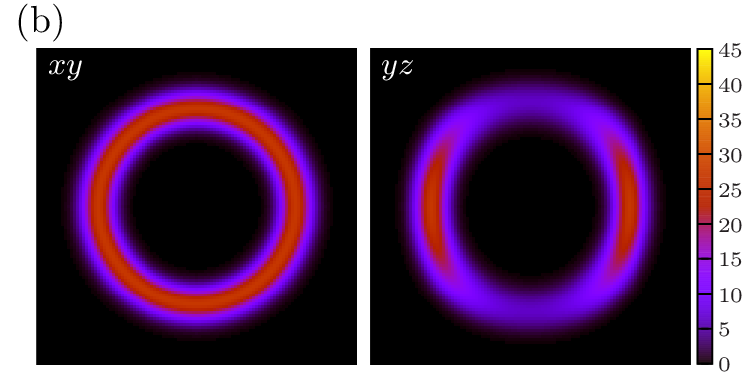}
\caption{\label{fig:gs-ddi}
Spherical shell-shaped trapping potential with frequency $\omega=2\pi\times 100$~Hz and radial shift $r_0=3.0~\mu$m. (a) Only contact interactions. (b) Both contact and dipolar interactions with $\epsilon_{\rm dd}=1.11$. The dipoles have magnetic moment $\mu=10 \,\muB$ and lie parallel to the $z$-axis. The left (right) panel of each case represents the 2D contour plot of the density in the $xy$ ($yz$) plane for a spherical shell-shaped BEC. Each plane shows a view of 16~$\mu$m $\times$ 16~$\mu$m, and the color scale corresponds to the density, which ranges from 0 (black) to 45 $\mathrm{\mu m}^{-3}$ (yellow). All the figures presented in this paper use the same color scale and numerical grid.
}
\end{figure}

For a spherical shell-shaped confining potential (\ref{eq-pot}), we show in figure~\ref{fig:gs-ddi} the 2D contour plot of the density in the $xy$ ($yz$) plane in the left (right) panel. Figure~\ref{fig:gs-ddi}(a) corresponds to a pure contact interacting BEC and figure~\ref{fig:gs-ddi}(b) to a dipolar BEC with the dipoles aligned along the $z$-axis. Due to the confinement, the condensate has a hollow core and is shell-shaped. The density distribution with only contact interactions---see figure~\ref{fig:gs-ddi}(a)---is entirely isotropic, while the addition of dipolar interactions produces a density accumulation around the equatorial region of the bubble---see right panel of figure~\ref{fig:gs-ddi}(b). In the equator, the dipoles lay mainly in a head-to-tail configuration, and the resulting interaction is attractive. In the polar regions, the dipoles sit instead side by side, which gives a neat repulsive interaction.

This anisotropic effect of the dipolar interaction was already shown in toroidal condensates \cite{Abad2010} and, more recently, in spherical shell-shaped potentials \cite{Adhikari2012}. Note that, despite the contour plot in the $xy$ plane is almost the same with and without dipolar interactions---see left panels on figure~\ref{fig:gs-ddi}(a) and (b)---, the maximum value of the density is higher in the presence of dipolar interactions than without them. The asymmetry in the density contour plot enhances as the relative strength between dipolar and contact interaction, $\epsilon_{\mathrm{dd}}$, increases \cite{Adhikari2012,Abad2010}.


\subsection{Gravitational sag}

The effect of gravity can be accounted for by including an additional potential term in the GPE (\ref{GPE}), the gravitational sag potential $V_g$~\cite{Sun2018}. To investigate the anisotropic effects of the dipolar interaction, we consider a general case in which the direction of gravity is not aligned with any of the axes of the trap but lies in the $xz$ plane. The gravitational sag potential reads
    \begin{equation}
        V_g(\vec{r}) = mg\left( x\sin\theta_0 + z\cos\theta_0\right),
    \end{equation}
where $\theta_0$ is the angle between the gravity direction and the $z$-axis.
In the particular case where the gravity and the $z$-axis are aligned ($\theta_0=0$), the gravitational sag $V_g(z) = mgz$ is equivalent to adding a vertical displacement to the trap's center~\cite{Jezek2004}.

Here we investigate the effects of gravity in the same spherical shell-shaped trapping potentials as in the previous subsection, both in contact-interacting and in dipolar BECs. We restrain our study to small strengths of gravity---with \emph{small} we mean larger than microgravity but smaller than Earth's gravity---since the terrestrial gravity destroys shell-shaped geometries~\cite{Lundblad2019}.

\subsubsection{Gravity aligned with the z-axis.}

\begin{figure}[t]
\centering
    \includegraphics[width = 0.51\linewidth]{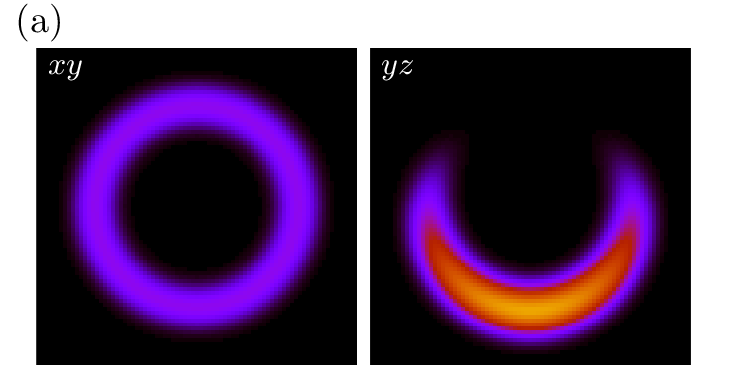}\hspace{-1.1em}
    \includegraphics[width = 0.51\linewidth]{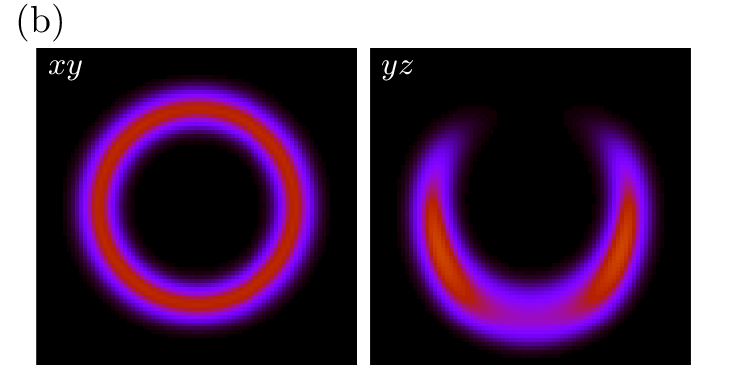}
\caption{\label{fig:gs-grv}
Same as figure~\ref{fig:gs-ddi} but in the presence of a small gravity of strength $0.005 \,\gE$ and parallel to the $z$-axis, with $\gE=9.8~\mathrm{m}\,\mathrm{s}^{-2}$ the terrestrial gravity. (a) Contact interactions only. (b) Contact and dipolar interactions, where dipoles have magnetic moment $\mu=10.0\,\muB$ and are aligned with the $z$-axis. The color scale and box size are the same as in figure~\ref{fig:gs-ddi}.
}
\end{figure}

We start by considering a gravity aligned with the $z$-axis and, for dipolar BECs, parallel with the polarization direction. Figure~\ref{fig:gs-grv} depicts the numerical results---see figure~\ref{fig:gs-ddi} for comparison without gravity. As we can see in the right panel ($yz$ plane) of the contact interacting case---figure~\ref{fig:gs-grv}(a)---, the atoms fall to the bottom of the shell-shaped potential. The density distribution in the $xz$ plane is the same as in the $yz$ plane due to the axial symmetry of the system: confinement, gravity, and polarization. The distortion of the trap, which results in a partially filled shell, is a clear signature of gravitational sag~\cite{frye2021}.

In the presence of dipolar interactions---figure~\ref{fig:gs-grv}(b)---, the interplay between their anisotropic character, the confining potential, and the gravitational sag leads to a partially filled shell, like in the contact-interacting case, but with a density depletion in the south region. As discussed in the situation with no gravity, the repulsive interaction between two parallel dipoles produces a significant reduction of the density in the bottom of the condensate---see the right panel of figure~\ref{fig:gs-grv}(b). The maximum density band lies slightly below the equatorial region, depending on the balance between the gravity and the dipolar moment of the atoms.

\subsubsection{Misaligned gravity.}

\begin{figure}[t]
\centering
\includegraphics[width = 0.51\linewidth]{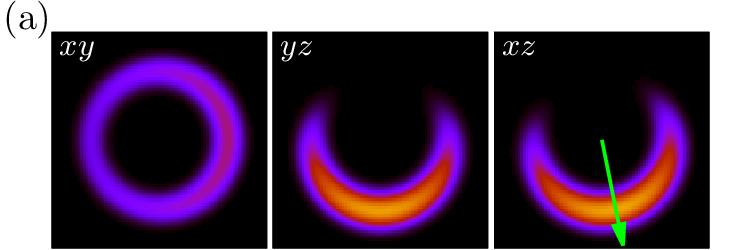}\hspace{-1.1em}
\includegraphics[width = 0.51\linewidth]{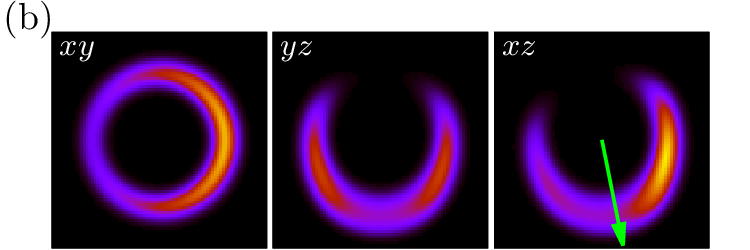}
\caption{\label{fig:gs-theta}
Same as figure~\ref{fig:gs-grv} but with an angle $\theta_0=-0.2$~rad between the gravity and the $z$-axis---such that gravity lies in the $xz$ plane. (a) Contact interactions only. (b) Dipolar interactions polarized along the $z$-axis. The green arrow in the $xz$ plane shows the direction of gravity.}
\end{figure}

We now explore a more general situation where the gravity and polarization direction are not aligned. Instead, gravity forms an angle $\theta_0$ with the $z$-axis and lies in the $xz$ plane. In figure~\ref{fig:gs-theta}, we show the 2D contour density plots in the three planes: $xy$ (left), $yz$ (middle), and $xz$ (right). Figure~\ref{fig:gs-theta}(a) corresponds to a pure contact interacting BEC, and figure~\ref{fig:gs-theta}(b) to a dipolar one.
For a contact interacting BEC, the 2D contour plot in the $yz$  plane remains almost unaltered as compared to figure~\ref{fig:gs-grv}(a), but the density's maximum in the $xz$ plane tilts in the direction of gravity---marked with a green arrow in the right panels of figure~\ref{fig:gs-theta}(a) and (b).
As one can see in the $xy$ plane, this tilting also produces an accumulation of particles in the right part of the bottom region of the shell.

The situation becomes more amusing for dipolar BECs, though, since the polarization axis fixes a privileged direction that breaks the symmetry when the gravity and the dipoles are not aligned. As a result, the density configurations in the $xz$ and $yz$ planes are now different from the contact-interacting case, as shown in figure~\ref{fig:gs-theta}(b). The density contour plot in the $yz$ plane is also similar to the density configuration when the gravity is parallel to the $z$-axis---see the right panel of figure~\ref{fig:gs-grv}(b). However, changes in the density in the $xz$ plane are more significant now: the maximum of the density lies in the right lobe of the shell and at a larger tilting angle compared to the direction of gravity. Within this region, the dipoles mainly lie head-to-tail, which results in an attractive interaction, whereas in the bottom of the shell (south pole), the atoms sit side by side, and hence the neat interaction is repulsive.

\begin{figure}[t]
	\centering
	\includegraphics[width = 0.5\linewidth]{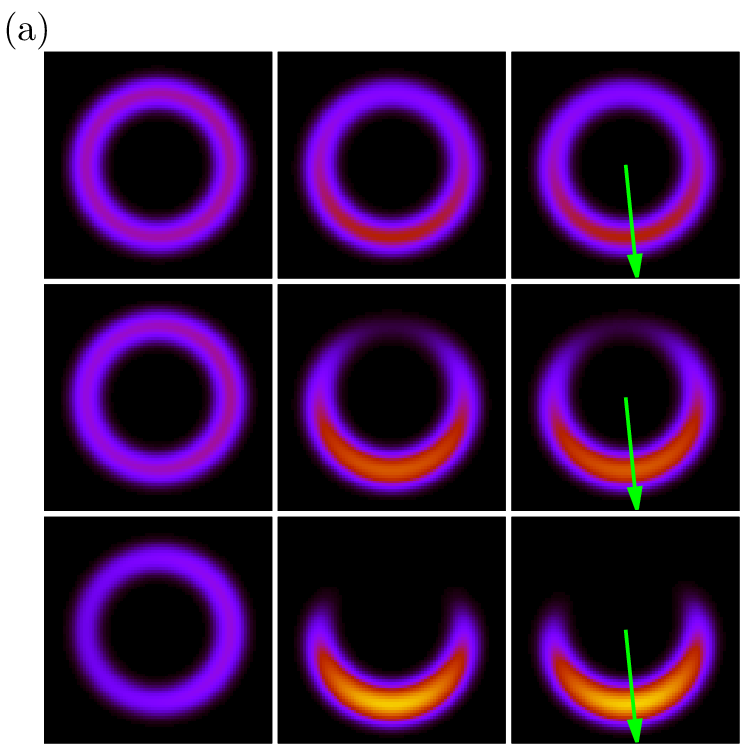}\hspace{-0.3em}
	\includegraphics[width = 0.5\linewidth]{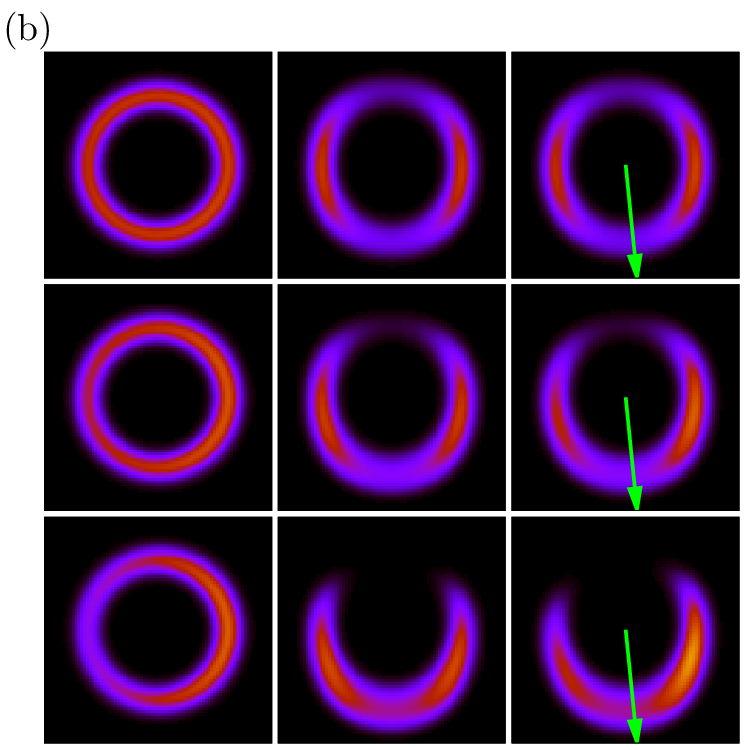}
	\caption{\label{fig:gs-compared}
	Same as figure~\ref{fig:gs-theta} with an angle $\theta_0=-0.1$~rad between the gravity and the $z$-axis. (a) Contact interactions only. (b) Dipolar interactions, with dipoles polarized along the $z$-axis. The strength of gravity for each case, from top to bottom, is $0.001\,\gE$, $0.003\,\gE$, and $0.006\,\gE$.
	}
\end{figure}

It is interesting to stress that this symmetry-breaking phenomenon, shown in the $xz$ plane, is produced by the anisotropic character of the dipolar interactions and depends both on the tilting angle $\theta_0$ and on the strength of gravity. In figure~\ref{fig:gs-compared}, we show the 2D contour plots of the density in the three planes ($xy$, $yz$ and $xz$) for a condensate with the same shell-shaped potential and for different small values of the gravity ($0.001 \leq g/\gE \leq 0.007$), tilted an angle $\theta_0=-0.1$~rad from the $z$-axis. Figure~\ref{fig:gs-compared}(a) corresponds to the numerical results for a contact interacting BEC, and figure~\ref{fig:gs-compared}(b) to a dipolar condensate. For small values of the strength of gravity (below $0.003 \,\gE$), the condensate forms a {\FS} with a higher density on the bottom. For slightly larger values (above $0.004 \, \gE$), the system is no longer a full shell due to the sag effect of the gravity; there are practically no atoms at the top of the trap, and the shape of the condensate is a hollow {\HS}. When we include dipolar interactions, their anisotropic character counterbalances the effect of gravity. As a result, the hole that appears at the top of the shell is small compared with the contact interacting case.


\section{Dynamics of small oscillations}\label{sec:dynamics}

In this section, we investigate the dynamical response of the system in the regime of small oscillations. To this aim, we trigger the dynamics by an instantaneous change in the tilting angle of gravity or its strength. We obtain the real-time evolution of the system by numerically solving the GPE (\ref{GPE}).

In the first scenario---subsection~\ref{subsec:dyn1}---, we consider gravity is initially tilted forming a small angle $\theta_0$ with the $z$-axis but contained in the $xz$ plane, and then it is suddenly aligned with the $z$-axis at $t=0$. In the second scenario---subsection~\ref{subsec:dyn2}---, the gravity is parallel with the $z$-axis ($\theta_0=0$), and we analyze the dynamics when slightly changing its strength from $g_0$ to $g$. To avoid large oscillations and complicated dynamics, we constrain our study to small variations.
Table~\ref{tab:dyn_cases} provides a summary of all the particular cases discussed in this section.

\begin{table}
\caption{\label{tab:dyn_cases}
Summary of numerical frequencies obtained from the oscillation of the center of mass for the particular cases studied in subsections~\ref{subsec:dyn1} and~\ref{subsec:dyn2}. We indicate the angle---with the $z$-axis---and strength of gravity and which of them is changed to trigger the dynamics. For each situation, we give the frequency for the BEC with only contact interactions (\textsc{ci}) and with both contact and dipolar interactions (\textsc{ddi}), and we indicate if the shape of the ground state is a {\HS} or a {\FS}.}
\begin{indented}
\item[]\begin{tabular}{@{}lccccc}
\br
&& \centre{2}{gravity} & \centre{2}{frequency (Hz)}\\
\ns
&&\crule{2}&\crule{2}\\
&& angle (rad) & strength ($\gE$) & \textsc{ci} & \textsc{ddi} \\
\mr
Variations in & \HS &
    $-0.1 \rightarrow 0.0$
    & $0.005$ &
    15.8 & 10.7\\
the angle$^{\rm (a)}$ & \FS &
    $-0.1 \rightarrow 0.0$
    & $0.002$ &
    15.9 & 10.6\\
\vspace{-1em}\\
Variations in & \HS &
    $0.0$
    & $0.005 \rightarrow 0.006$ &
    24.7 & 25.1\\
the strength$^{\rm (b)}$ & \FS &
    $0.0$
    & $0.003 \rightarrow 0.002$ &
    16.1 & 22.2\\
\br
\end{tabular}
\item[] $^{\rm (a)}$ Frequency calculated from the oscillation of $\langle{x(t)}\rangle$. See subsection~\ref{subsec:dyn1}.
\item[] $^{\rm (b)}$ Frequency calculated from the oscillation of $\langle{z(t)}\rangle$. See subsection~\ref{subsec:dyn2}.
\end{indented}
\end{table}

\subsection{Variations in the orientation of gravity}\label{subsec:dyn1}

For all the results presented here, we consider $\theta_0=-0.1$~rad. We have checked that, for a given strength of gravity, the dynamics are the same independently of the sign and value of the initial tilting angle as long as such angle is small. We open this subsection with a detailed study of two particular cases---one with $g>0.004\,\gE$ and the other with $g<0.004\,\gE$---to see how the shape of the ground state affects the dynamics.
The results we show for discussion are the oscillations of the center of mass---figure ~\ref{fig:evolution}, see numerical frequencies in table~\ref{tab:dyn_cases}---and some snapshots of the density during the first period of the evolution---figure~\ref{fig:dyn-angle}.
Lastly, we analyze how the oscillation frequency depends on the strength of gravity---see figure~\ref{fig:frequencies_angle}.

\subsubsection{Particular cases.}

\begin{figure}[t]
\centering
\includegraphics[width = 0.45\linewidth]{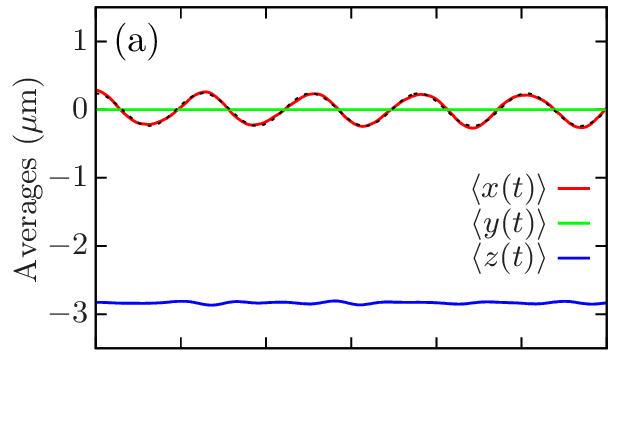}\hspace{-2em}%
\includegraphics[width = 0.45\linewidth]{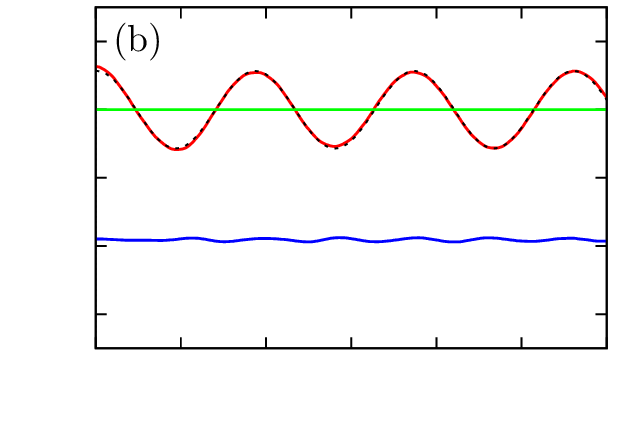}\vspace{-1em}
\includegraphics[width = 0.45\linewidth]{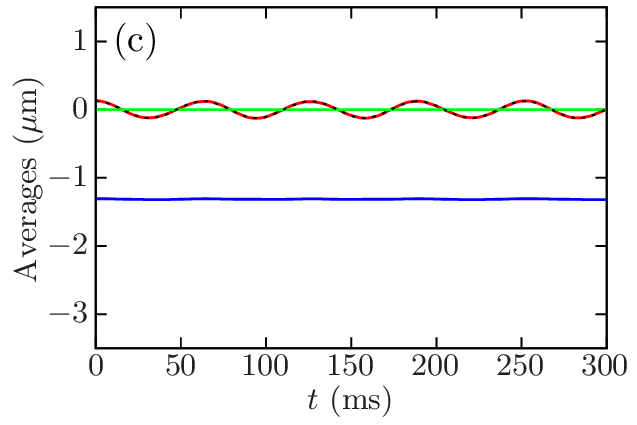}\hspace{-2em}%
\includegraphics[width = 0.45\linewidth]{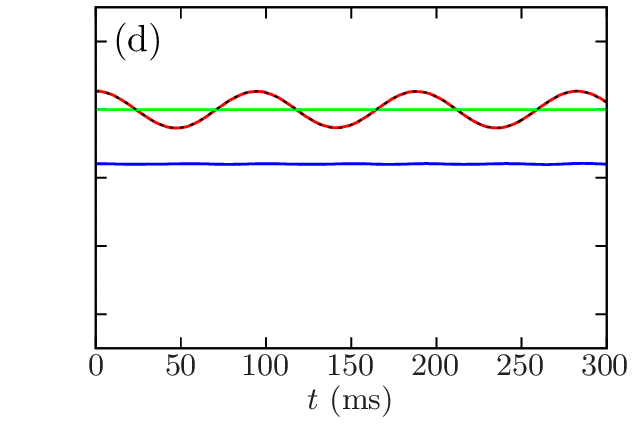}
\caption{
Numerical evolution of the coordinates of the center of mass. Gravity is initially at an angle $\theta_0=-0.1$~rad with the $z$-axis. We consider the two possible static regimes described in the text: a {\HS} for $g=0.005\,\gE$ with (a) contact and (b) dipolar interactions, and a {\FS} for $g=0.002\,\gE$ with (c) contact and (d) dipolar interactions. The black dashed lines correspond to the sinusoidal fits of the numerical results for $\langle x(t) \rangle$, from where the frequencies of oscillation are obtained: (a) $15.8$~Hz, (b) $10.7$~Hz, (c) $15.9$~Hz, and (d) $10.6$~Hz. See summary of numerical frequencies in table~\ref{tab:dyn_cases}.  } 
\label{fig:evolution}
\end{figure}
\begin{figure}[t]
\centering
\includegraphics[width = 0.24\linewidth]{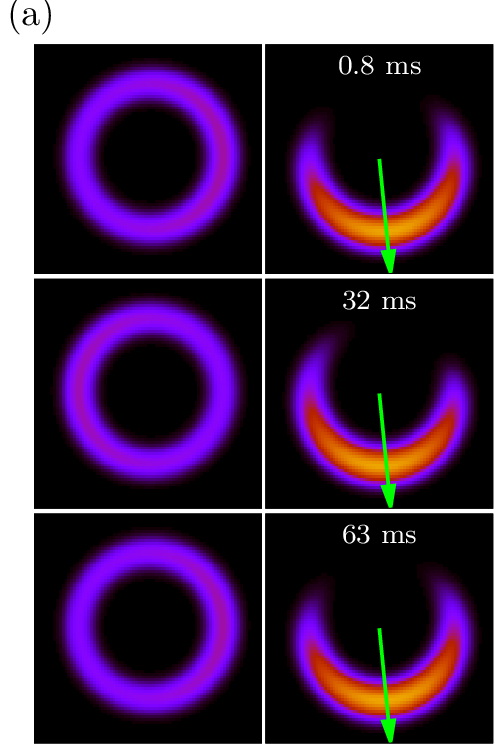}
\includegraphics[width = 0.24\linewidth]{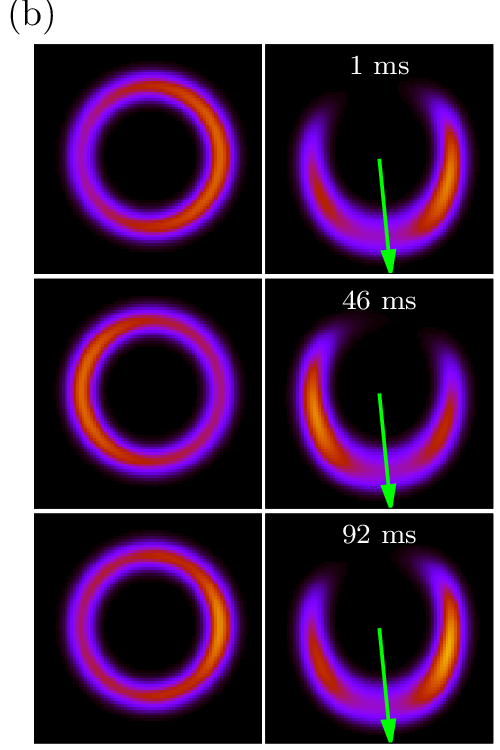}
\hfill
\includegraphics[width = 0.24\linewidth]{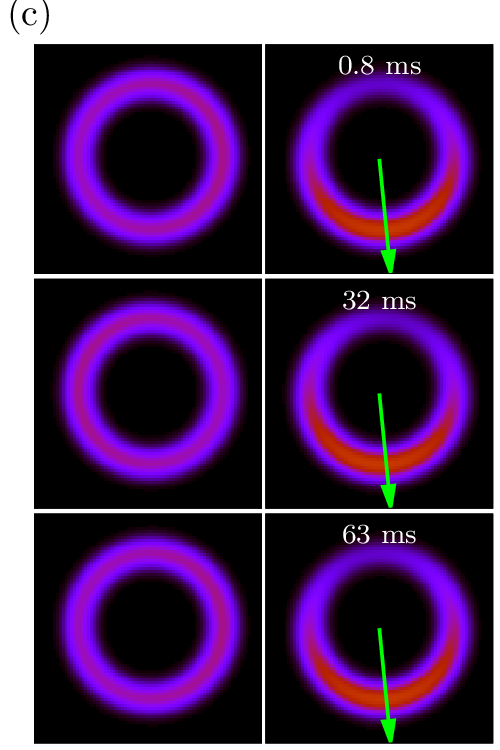}
\includegraphics[width = 0.24\linewidth]{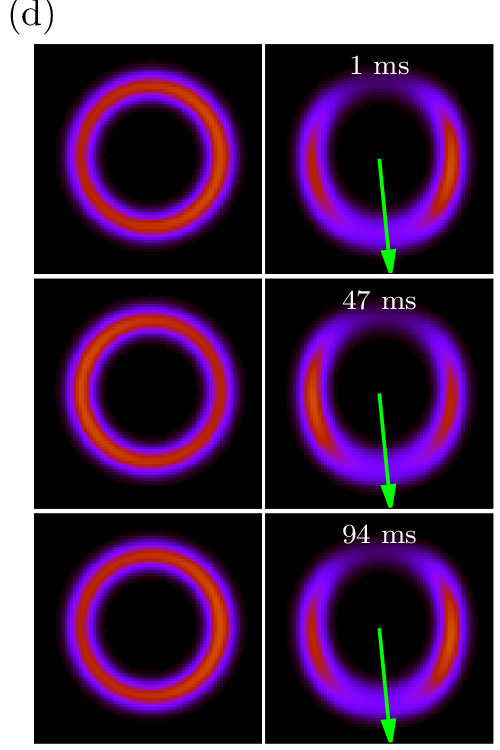}
\caption{\label{fig:dyn-angle}
Snapshots of the 2D contour plots of the density in the $xy$ (left panels) and $xz$ (right panels) planes at different times of the evolution. We discarded the $yz$ planes because the densities remain unchanged, as the variation in gravity is constrained to the $xz$ plane. The initial tilt of the gravity is $\theta_0=-0.1$~rad, and we study the same two situations as in figure~\ref{fig:evolution}: a {\HS} for $g=0.005\,\gE$ with either (a) only contact interactions or (b) both contact and dipolar interactions, and a {\FS} for $g=0.002\,\gE$ with (c) only contact and (d) also dipolar interactions. In both dipolar cases, the dipoles have magnetic moment $\mu=10\,\muB$. The green arrow shows the initial direction of gravity, which is later aligned with the $z$-axis to start the dynamics. See summary of cases in table~\ref{tab:dyn_cases}.
}
\end{figure}


We start with a gravity of strength $g=0.005\,\gE$ such that the system resembles a {\HS}, as we discussed in section~\ref{sec:ground}. First, we consider a shell-shaped BEC with only contact interactions. In figure~\ref{fig:evolution}(a), we present the time evolution of the coordinates of the center of mass: $\langle{x(t)}\rangle$, $\langle{y(t)}\rangle$, and $\langle{z(t)}\rangle$. Since we prepare the system with a slight misalignment of the gravity, the sudden alignment with the $z$-axis forces the system to bounce back and forth in the $xz$ plane around the new equilibrium position, the $z$-axis. This behavior appears as a sinusoidal-like oscillation of $\langle x(t) \rangle$ as a function of time, while the other coordinates remain almost unaltered. The sinusoidal fit of the numerical evolution of $\langle x (t) \rangle$ gives a frequency of $15.8$~Hz---we checked that the frequency of oscillation is close to this value when the initial angle $|\theta_0|$ is approximately below $0.15$~rad. Figure~\ref{fig:dyn-angle}(a) displays a few snapshots---the times shown cover a whole period---of the 2D contour plots of the density in the two planes where the oscillations are observable, $xy$ and $xz$. Since the shell-shaped BEC is 3D, the oscillatory behavior of $\langle x(t) \rangle$---see figure~\ref{fig:evolution}(a)---produces symmetric rearrangements of the density in the other directions, as shown in the $xy$ plane of figure~\ref{fig:dyn-angle}(a).

Figure~\ref{fig:evolution}(b) shows the numerical evolution of the center of mass for a dipolar condensate with an initial tilting angle of gravity $\theta_0=-0.1$~rad. As we discussed before---see section~\ref{sec:ground} and figure~\ref{fig:gs-compared}(b)---, the filled region of the shell-shaped potential appears at a larger tilting angle in a dipolar condensate than in a contact interacting one. This feature of the anisotropy of the dipolar interactions produces a larger amplitude of the oscillations of $\langle x(t) \rangle$ in dipolar BECs. The sinusoidal fit of the numerical evolution of $\langle x(t) \rangle$ gives a frequency of $10.7$~Hz; as in the contact interacting case, the other components of the center of mass of the system, $\langle{y(t)\rangle}$ and $\langle{z(t)\rangle}$, show practically no variations. When the gravity is suddenly aligned, the system oscillates around the $z$-axis as expected. However, unlike in the contact interacting case, the atoms do not pass over the south pole of the {\HS}, where the neat dipolar interaction is repulsive: their movement is instead constrained to the high-density band that appears below the equatorial region. One can see this behavior in figure~\ref{fig:dyn-angle}(b), which shows a few snapshots covering one period in the $xy$ and $xz$ planes.


Lastly, we study the situation where the gravity is small enough---in particular, $g=0.002\,\gE$---that the BEC still retains its {\FS} shape. From figure~\ref{fig:evolution}(c) and (d), we can see that the oscillations of $\langle{x(t)}\rangle$ are broader and slower in the dipolar BEC than in the contact interacting one, as in the previous case. The frequencies of oscillation we obtain from the fit are $15.9$~Hz (contact BEC) and $10.6$~Hz (dipolar BEC), which resemble those from the previous case. If we compare the oscillations---figure~\ref{fig:evolution}(c) and (d)---with those obtained for a heavier gravity---figure~\ref{fig:evolution}(a) and (b)---, we observe that the frequencies are similar in both the contact and the dipolar BECs, but the amplitudes of the oscillations are much lower now. From the snapshots of the density shown in figure~\ref{fig:dyn-angle}(c) and (d), we can see that in the case of a smaller gravity, as expected, the atoms can move around the whole shell---not just the lower part---, which could explain why the oscillations of the center of mass in the $x$ direction are more restricted. We will explore in more detail the effect of gravity in the dynamics in the following subsection.

\subsubsection{The role of gravity.}

\begin{figure}[t]
\centering
\includegraphics[width=0.14\linewidth]{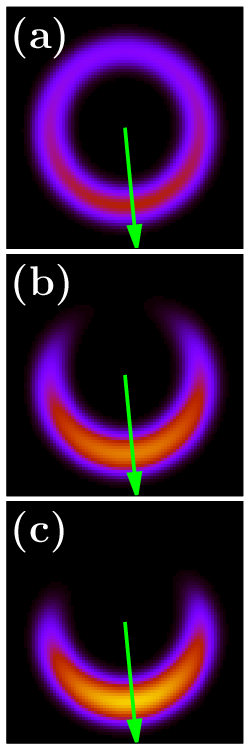}\hspace{1em}%
\includegraphics[width=0.6\linewidth]{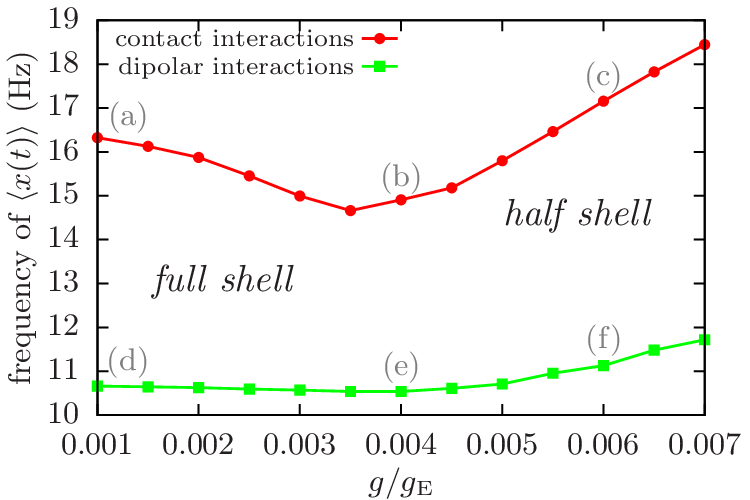}\hspace{1em}%
\includegraphics[width=0.14\linewidth]{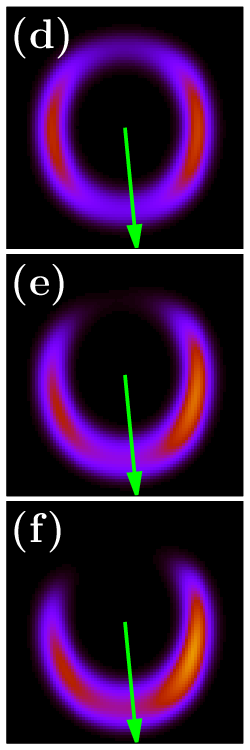}
\caption{Oscillation frequency of $\langle{x}(t)\rangle$ as a function of the gravity with contact (red) and dipolar (green) interactions, where the gravity has an initial tilting angle $\theta_0=-0.1$~rad. We obtain the oscillation frequency by fitting a sinusoidal function to the numerical data. The panels on both sides show the 2D contour plots of the initial density in the $xz$ plane---which contains gravity---for the different values of the gravity labeled from (a) to (f), both for the contact interacting BEC (left panels) and the dipolar one (right panels). The green arrow, as in the previous figures, marks the direction of gravity. Lines between data points are added to guide the eye.}
\label{fig:frequencies_angle}
\end{figure}

Here, we study the dynamics of small oscillations due to variations in the tilting angle---initially $\theta_0=-0.1$~rad in all the cases---for different strengths of gravity. In figure~\ref{fig:frequencies_angle}, we show the oscillation frequency of the $x$ coordinate of the center of mass, $\langle{x(t)}\rangle$, as a function of the strength of gravity. We consider BECs with both contact and dipolar interactions. As figure~\ref{fig:frequencies_angle} shows, the oscillation frequency depends on the strength of gravity, and two different behaviors arise: first, starting from the lowest gravity, the frequency decreases as $g$ increases until it reaches a particular value (between $0.003\,\gE$ and $0.004\,\gE$); then it increases again. These two behaviors are related to the two distinct shapes that can be observed in the ground states of the system for different values of gravity, as is shown in figure~\ref{fig:gs-compared} and discussed in section~\ref{sec:ground}: when the strength of gravity is small, the ground state of the system is a {\FS}, while for heavier values of gravity it resembles a {\HS}.

At small strengths of gravity, the condensate is a {\FS} with a higher density at the
bottom of the trap. Then, an increase in gravity drags more atoms to the bottom of the trap, which leads to a decrease in the oscillation frequency. When considering dipolar interactions, though, their anisotropic nature compensates for the effect of gravity; as a result, the oscillation frequency becomes almost invariant to small changes in the strength of gravity.

On the other hand, at larger values of $g$, the system is no longer a {\FS} but a {\HS}, and the oscillation frequency increases as the strength of gravity does---as one would expect from a pendulum, where the frequency grows proportionally to $\sqrt{g}$. For a mathematical pendulum with a fixed length, a change of gravity (from $g_1$ to $g_2$) is related to the change in the oscillation frequency (from $\omega_1$ to $\omega_2$) as $\omega_1/\omega_2=\sqrt{g_1/g_2}$. Using the numerical results from figure~\ref{fig:frequencies_angle}, we checked that this relation holds when $g>0.004\,\gE$ for the contact interacting case and when $g>0.005\,\gE$ for the dipolar interacting one.


\subsection{Variations in the strength of gravity}\label{subsec:dyn2}

In the previous subsection---\ref{subsec:dyn1}---, we discussed the dynamics due to variations in the angle of gravity. Here we fix the angle of gravity with the $z$-axis to zero ($\theta_0=0$) and study the system's response to variations in the strength of gravity. As before, we constrain our study to small oscillations, which now translates to small variations in the strength of gravity. We start by preparing the system under a gravity $g_0$ aligned with the $z$-axis, and then, at $t=0$, we change $g_0$ to $g$.

In the first part of this subsection, we study in detail two cases: first, when $g, g_0 > 0.004\,\gE$, so the corresponding ground states resemble a {\HS}, as discussed in section~\ref{sec:ground}; then, we set $g, g_0 < 0.004\,\gE$, with both values of gravity laying in the regime where the system is still a {\FS}.
See table~\ref{tab:dyn_cases} for a summary of the numerical frequencies obtained and figure~\ref{fig:dyn-gravity} for some snapshots of the evolution.
For these cases, we choose a large change in gravity ($|g-g_0|=0.001\,\gE$) to see the system's dynamics well.
In the second part, we determine for which variations of gravity the oscillations can be considered small, and then we fix the value of the variation ($|g-g_0|=0.0001\,\gE$) and compare the frequencies of oscillation obtained for different values of the final gravity $g$---see figure~\ref{fig:frequencies_strength}.

\subsubsection{Particular cases.}

\begin{figure}[t]
\centering
\includegraphics[width = 0.24\linewidth]{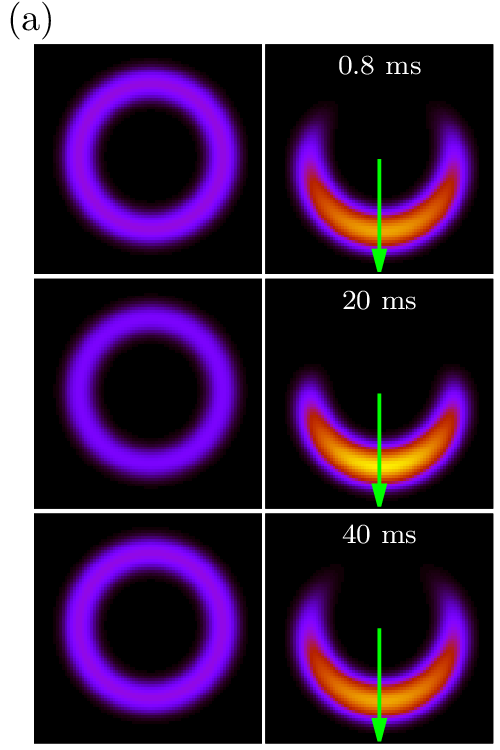}
\includegraphics[width = 0.24\linewidth]{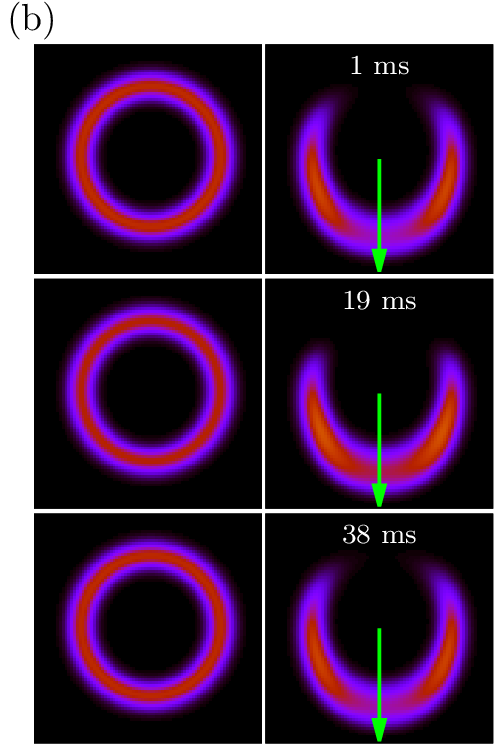}
\hfill%
\includegraphics[width = 0.24\linewidth]{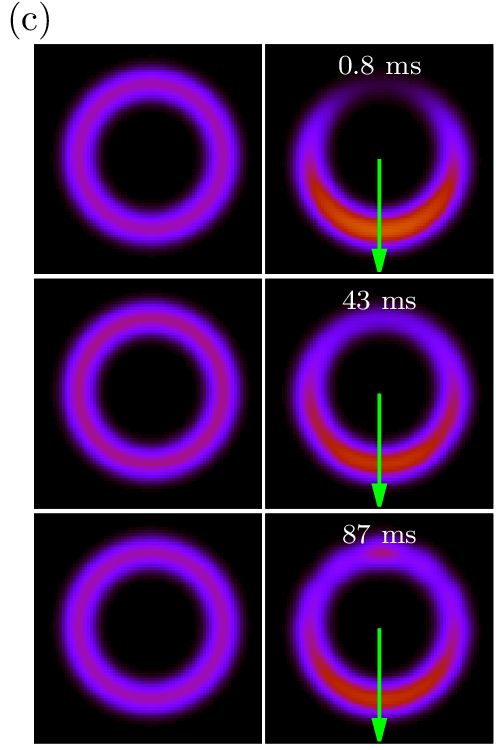}
\includegraphics[width = 0.24\linewidth]{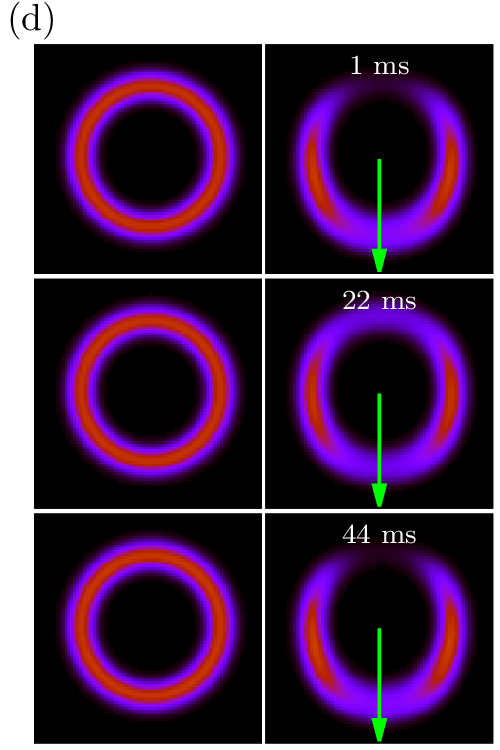}
\caption{\label{fig:dyn-gravity}
Snapshots of the 2D contour plots of the density in the $xy$ (left) and $xz$ (right) planes at different times of the evolution. Gravity is parallel to the $z$-axis, and its strength is varied from $g_0$ to $g$ at $t=0$. Since the densities in the $xz$ and $yz$ planes are equivalent, we discarded the
$yz$ planes. First case, {\HS}: $g_0=0.005\,\gE$ and $g=0.006\,\gE$, with (a) contact interactions only and (b) contact and dipolar interactions. Second case, {\FS}: $g_0=0.003\,\gE$ and $g=0.002\,\gE$, with (c) contact interactions and (d) dipolar interactions. The dipole moment, $\mu=10\,\muB$, is the same for all the cases with dipolar interactions. See summary of cases in table~\ref{tab:dyn_cases}.
}
\end{figure}

In the first case of our study, the initial strength of gravity is $g_0=0.005\,\gE$, and the evolution starts when we abruptly increase it to $g=0.006\,\gE$. Within these values of the gravity, the ground state of the system resembles a {\HS}, as we already mentioned---see the last row in figure~\ref{fig:gs-compared}. In figure~\ref{fig:dyn-gravity}(a) and (b), we plot the densities at different times---the snapshots cover a whole period of the oscillation---to show the dynamics of both the contact and dipolar cases. In the contact-interacting case---figure~\ref{fig:dyn-gravity}(a)---, the atoms are mainly located at the south pole of the shell, occupying a region that shrinks and grows periodically due to the increase in gravity. This behavior resembles a spring that oscillates vertically. Here, though, the movement of the atoms is confined to the surface of the shell. In the dipolar case---figure~\ref{fig:dyn-gravity}(b)---, instead, the band of maximum density appears below the equatorial region. Then, the sudden change in gravity causes this band to oscillate along the $z$ direction. Since the gravity is parallel to the $z$-axis, we study the oscillation frequency of the $z$ coordinate of the center of mass through a sinusoidal fit to the numerical results for $\langle{z(t)}\rangle$. We obtain a frequency of $24.7$~Hz for the contact-interacting BEC and $25.1$~Hz for the dipolar one. Unlike in subsection~\ref{subsec:dyn1}, here we find that both frequencies are similar.

For the second case, where the gravity is small enough that the system has the shape of a {\FS}, we decrease the initial gravity $g_0=0.003\,\gE$ to $g=0.002\,\gE$. The dynamics are very similar to the previous case, as figure~\ref{fig:dyn-gravity}(c) and (d) shows. In this case, however, the oscillation frequencies of $\langle{z(t)}\rangle$ for the contact and dipolar BECs are more different; in particular, we find $16.1$~Hz for the contact-interacting BEC and $22.2$~Hz for the dipolar one.

The oscillations of the center of mass---in both frequency and amplitude---depend on the strength of gravity. These results match those found in subsection~\ref{subsec:dyn1}. In contact-interacting BECs, when the gravity is light---and the system is a {\FS}---, we find that the oscillations are slow and broad since the atoms can move around the whole shell. For a heavier gravity, the atoms drop to the bottom of the trap. Then, the amplitude of the oscillation decreases while its frequency increases. The differences found in dipolar BECs come from the anisotropic nature of the dipolar interactions, which counterbalances gravity. The band of maximum density is no longer at the south pole but below the equator. Therefore, compared to the contact-interacting case, the oscillations change much less when different strengths are studied.


\subsubsection{The role of gravity.}

Finally, we study the dynamics of small oscillations induced by a variation in the strength of gravity and how these results differ depending on whether the value of gravity is relatively small---and the ground state resembles a {\FS}---or large---when the system becomes a {\HS}.

To determine how much we can vary the gravity without having large oscillations, we consider the following argument. We can take, for instance, a case where the system resembles a {\HS}---when we expect it behaves like a pendulum---and then, for a given initial gravity $g_0$, calculate the oscillation frequency for different values of the final gravity. As we did previously, we obtain the frequency by fitting a sinusoidal function to the oscillation of the $z$ coordinate of the center of mass. In particular, we consider $g_0=0.005\,\gE$ and study some cases with final gravity $g/\gE\in [0.004,0.006]$. Then, plotting the frequency as a function of the variation of gravity $(g-g_0)/\gE$, we can see that the frequency grows somewhat linearly with the variation of gravity for values of the variation close to zero, while for larger values---in absolute value---, its behavior is no longer linear. This range of apparent linear behavior corresponds to $|g-g_0|<0.0003\,\gE$. Thus, we assume that the oscillations will be small as long as the gravity variation is below this threshold, and hence we choose $|g-g_0|=0.0001\,\gE$ to ensure small oscillations. For these values of $g-g_0$, the results we obtain for a given $g$ are equivalent---in frequency and amplitude---either if $g>g_0$ or $g<g_0$. Therefore, we define from now on $g_0$ such that $g_0=g+0.0001\,\gE$.

\begin{figure}[t]
\centering
\includegraphics[width=0.6\linewidth]{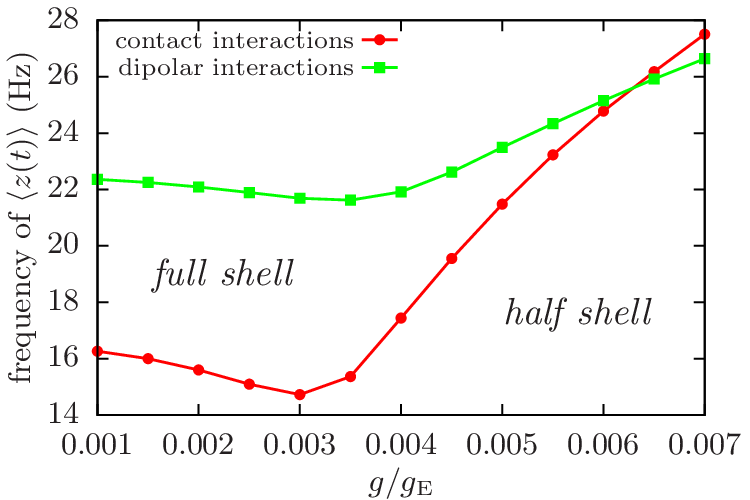}
\caption{
Oscillation frequency of $\langle{z}(t)\rangle$ as a function of the final gravity $g$, with initial gravity $g_0=g+0.0001\,\gE$. The red line corresponds to the case with only contact interactions, and the green line to the dipolar interacting BEC. As before, we obtain the oscillation frequency by fitting a sinusoidal function to the numerical data. Lines are added to guide the eye.}
\label{fig:frequencies_strength}
\end{figure}

In figure~\ref{fig:frequencies_strength}, we plot the frequency of oscillation of $\langle{z(t)}\rangle$ as a function of the final gravity $g$. The results resemble those from figure~\ref{fig:frequencies_angle}. The frequency increases with the final gravity for large values of the gravity---when the system is a {\HS} and behaves like a pendulum---, while it decreases with the final gravity for smaller values---when the system resembles a {\FS}. Since the dipolar interaction compensates for the gravity, the effect of the variation in strength is more noticeable in the contact-interacting BEC than in the dipolar BEC---as in subsection~\ref{subsec:dyn1}---for small final gravities.

Comparing the results obtained for the contact-interacting BEC either with changes in strength---see figure~\ref{fig:frequencies_strength}, red line---and orientation---see figure~\ref{fig:frequencies_angle}, red line---, we can see that the frequencies lie within a similar range of values in both cases. The frequencies we obtain now for the dipolar BEC, however, are faster. This increase in frequency is an effect of the anisotropy of the dipolar interactions. In the first scenario---see figure~\ref{fig:frequencies_angle}, green line---the center of mass moves mainly along the $x$-axis, and all the dipoles point towards the $z$-axis. Then, an atom that moves in that direction feels a net repulsive interaction from its neighbors, which reduces the frequency of oscillation. In the second scenario---see figure~\ref{fig:frequencies_strength}, green line---the center of mass moves instead around the $z$-axis. Since the resulting interaction between dipoles along the direction of motion is attractive and twice as large as in the previous case, the frequency of the oscillation is much larger.


\section{Summary and outlook}\label{sec:conclusions}

In the present work, we have studied the statics and dynamics of shell-shaped condensates with contact and dipolar interactions in the presence of a small gravity. We have constrained our study to gravity values above microgravity---and thus non-negligible---and below terrestrial gravity, which destroys shells.

First of all, we have analyzed the ground states of the system in three cases: without gravitational sag, with gravity parallel to the $z$-axis---which is the polarization direction we considered for dipolar BECs---, and with a small gravity misaligned with the $z$-axis and contained in the $xz$ plane. We have discussed the effect of the dipolar interactions in either of the three cases. We have shown that since the dipolar interaction adds a privileged direction to the one already defined by gravity, the resulting shells with gravitational sag and dipolar interactions present an attractive configuration to study misalignments and perturbations in the gravitational sag.
Next, we have done a more general analysis of the ground states to examine the effect of gravity's strength. Observing the shape that the system displays, we have defined two regimes: a {\FS} for small gravities and a {\HS} for larger gravities. These two regimes play a relevant role in the dynamical behavior of the system.

Later, we have studied the dynamics of small oscillations due to changes in the orientation and strength of gravity. For each of those two scenarios, we have studied two particular cases---comparing the {\FS} and {\HS} regimes---, and we have analyzed, more generally, the effect that gravity has in the behavior of the oscillations when the variation---in angle or strength---is fixed and very small. With this, we have seen how the two static regimes translate into two distinct dynamical behaviors: the oscillation frequency increases with gravity for large values of gravity ({\HS}) while it decreases for smaller values ({\FS}).
Additionally, we have compared the results obtained for contact-interacting BECs with those obtained for their dipolar-interacting counterparts. We have discussed that dynamics due to changes in angle or strength are equivalent in the contact BECs, but the dynamical behavior differs in dipolar BECs due to the anisotropic nature of their interactions.

The atomic cloud in this system is not only sensitive to changes in its orientation, but it is also sensitive to small gravitational variations, either in its direction or strength. This result could pave the way to the experimental realization of a gravity or accelerometer sensor intended for small gravity conditions. Monitoring gravity and its changes from space in satellite missions---see~\cite{Migliaccio2019} and references therein---is another possible application of this system. To conclude, we want to point out the experimental feasibility of the proposed system. We have used values for the experimental parameters that are currently available in laboratories.

Due to the complexity of the 3D dynamics of this system, a more exhaustive theoretical analysis is beyond the scope of this paper. However, we consider that
a new proposal of a gravity sensor with restricted low dimensional dynamics may provide a more analytical insight of the system. Discussing other configurations such as toroidal BECs under gravity conditions will be addressed elsewhere.

\ack
We thank Albert Roura for useful discussions.
This work has been supported by Grant No. FIS2017-87801-P (AEI (ES), FEDER (EU)),
by Grant No. PID2020-114626GB-I00
(Ministerio de Ciencia e Innovaci\'on)
and
by the European Union Regional Development Fund within the ERDF Operational Program of Catalunya (project QUASICAT/QuantumCat).
M. A. is supported by FPI Grant PRE2018-084091.

\section*{References}

\bibliographystyle{iopart-num}
\bibliography{bibliography}

\end{document}